\documentclass[aps,prd,onecolumn,groupedaddress,showpacs,nofootinbib,amssymb]{revtex4}
\usepackage[dvips]{graphicx}
\usepackage{amssymb}
\usepackage{amsmath}
\usepackage{graphicx,,color}
\usepackage{amsfonts}
\usepackage{bm}
\usepackage{cancel}
\usepackage{comment}
\newcommand\be{\begin{equation}}
\newcommand\ee{\end{equation}}

\allowdisplaybreaks[4]

\begin{document}
\tolerance=5000

\title{Inflation with Gauss-Bonnet and Chern-Simons higher-curvature-corrections in the view of GW170817}
\author{S.A.
Venikoudis,\,\thanks{venikoudis@gmail.com} F.P.
Fronimos\, \thanks{fotisfronimos@gmail.com}}
\affiliation{
 Department of Physics, Aristotle University of
Thessaloniki, Thessaloniki 54124,
Greece\\}

\tolerance=5000


\begin{abstract}
Inflationary era of our Universe can be characterized as semi-classical because it can be described in the context of four-dimensional Einstein's gravity involving quantum corrections. These string motivated corrections originate from  quantum theories of gravity such as superstring theories and include higher gravitational terms as, Gauss-Bonnet and Chern-Simons terms. In this paper we investigated inflationary phenomenology
coming from a scalar field, with quadratic curvature terms in the view of GW170817. Firstly, we derived the equations of motion, directly from the gravitational action. As a result, 
formed a system
of differential equations with respect to Hubble’s parameter and the inflaton field which was very
complicated and cannot be solved analytically, even in the minimal coupling case. Based on the observations from GW170817 event, which have shown that the speed of the primordial gravitational wave is equal to the speed of light, $c_\mathcal{T}^2=1$ in natural units, our
equations of motion where simplified
after applying the constraint $c_\mathcal{T}^2=1$, the slow-roll approximations and neglecting the string corrections. We described the dynamics of inflationary phenomenology and proved that theories with Gauss-Bonnet term can be compatible with recent observations. Also, the Chern-Simons term leads to asymmetric generation and evolution of the two circular polarization states of gravitational wave. Finally, viable inflationary models are presented, consistent with the observational constraints. The possibility of a blue tilted tensor spectral index is briefly investigated.

\end{abstract}
\maketitle

\section{Introduction}
Cosmology is currently at the stage in which the existence of the observational data needs to be appropriately explained
theoretically. The three most fascinating mysteries in modern Cosmology are, the nature of the dark matter, the dark energy issue for the late-time Universe and the primordial era.
Until now, based on the constraints from observations, we are at the stage of speculating and fitting models that may appropriately describe these in a
consistent way. Regarding to the primordial era, our approach is through a regime of classical gravity towards to the unknown era of quantum gravity which is believed to govern the small scale physics and unifies all the fundamental forces in nature. In between the classical gravity regime and the quantum gravity regime, it is believed that an era of abrupt accelerated with quasi-exponential rate expansion occurred, known
as inflationary era.
The effective Lagrangian of inflation is not specified by the data
at present time. Thus, although the inflationary era is can be considered as a classical
era of our Universe, which is described by a four dimensional
spacetime, it still is possible that the quantum era may have a direct
imprint on the effective Lagrangian of inflation.
Therefore, some of the most simple corrections of the inflationary effective Lagrangian may
be provided by higher curvature terms like, $f(R)$ gravity
\cite{Nojiri:2017ncd,Capozziello:2011et,Capozziello:2010zz,Nojiri:2006ri,Nojiri:2010wj,Olmo:2011uz}, Gauss-Bonnet corrections \cite{Kanti:2015pda,Yi:2018gse,Guo:2010jr,Jiang:2013gza,Guo:2009uk,DeLaurentis:2015fea,Fomin:2020hfh,Pozdeeva:2020apf,Yi:2018dhl,vandeBruck:2016xvt,Odintsov:2018zhw,Nozari:2017rta,Chakraborty:2018scm,Kawai:1999pw,vandeBruck:2017voa,Bakopoulos:2020dfg,Kleihaus:2019rbg,Bakopoulos:2019tvc,Kanti:1995vq,Bajardi:2019zzs} and also Chern-Simons corrections \cite{Nojiri:2020pqr,Nojiri:2019nar,Odintsov:2019evb,Odintsov:2019mlf,Alexander:2009tp,Qiao:2019hkz,Nishizawa:2018srh,Wagle:2018tyk,Yagi:2012vf,Yagi:2012ya,Molina:2010fb,Izaurieta:2009hz,Smith:2007jm,Konno:2009kg,Sopuerta:2009iy,Matschull:1999he,Haghani:2017yjk}.
Literature on inflationary cosmology with the existence of a combination of the quadratic gravitational terms is discussed in Ref. \cite{Kawai:2017kqt,Satoh:2007gn,Nair:2019iur,Satoh:2008ck,Satoh:2010ep,Antoniadis:1993jc}.

Recently,  the
LIGO-Virgo detectors observed a gravitational wave coming from the merging of two neutron stars \cite{GBM:2017lvd}. 
The interesting fact of the observation is that the gravitational waves arrived almost simultaneously
with the gamma rays emitted from the neutron stars merging. Thus, the
speed of the gravitational wave $c_\mathcal{T}$ is implied to be nearly equally to unity, where $c_\mathcal{T}^2$ in natural units, which is equal to the speed of light.
The detection of GW170817 event utterly changed the scenery in modern theoretical Cosmology, by excluding many modified theories of gravity because they predict primordial massive gravitons \cite{Ezquiaga:2017ekz}. Einstein-Gauss-Bonnet theories have a serious drawback according our previous reasoning. They predict a non-zero mass for gravitons during the primordial inflationary era. Until now, there is no mechanism in particle physics to explain why the graviton should change mass during different time periods of the Universe. According to our previous assumption, Einstein's gravitational theory with quadratic curvature corrections can be compatible with the GW170817 and produce gravitational waves with the speed of light as proved in our previous works\cite{Odintsov:2020sqy,Oikonomou:2020sij,Odintsov:2020xji,Oikonomou:2020oil,Odintsov:2020mkz,Oikonomou:2020tct,Venikoudis:2021irr}.
After imposing the condition $c_{\mathcal{T}}^2=1$, the coupling functions and the scalar potential of the theory are restricted and the theory can predict viable inflationary era according to the latest Planck data Ref.\cite{Akrami:2018odb}.

In this paper we shall investigate whether is possible to achieve viable
inflationary models with Gauss-Bonnet and Chern-Simons higher curvature corrections consistent with the latest Planck data. The addition of the Chern-Simons term in 
 the Lagrangian does not affect the background equations and the scalar spectral index of primordial curvature perturbations, it affects only the tensor perturbations of the theory. The Chern-Simons term leads to an additional constraint in our theory in order to be compatible with the GW170817. More specifically, the existence of the parity violating term leads to asymmetric generation and evolution of the two circular polarization states of gravitational
wave as proved in Ref. \cite{Hwang:2005hb}. Furthermore, due to the Chern-Simons term the generation of a blue-tilted becomes a plausible scenario which is investigated in the non-minimally coupled case for the sake of generality. 

The present paper is organized as follows: In section II the theoretical framework of slow-roll inflationary phenomenology in Einstein's gravity in the presence of Gauss-Bonnet and Chern-Simons higher curvature corrections is presented and moreover the impact of such terms in the slow-roll indices and in consequence the observed indices is investigated.
In section III several viable inflationary models, compatible with the latest Planck data, minimally and non-minimally coupled with the Ricci scalar are presented. 
Finally, the conclusions follow in the end of the paper.

\section{INFLATIONARY PHENOMENOLOGY WITH HIGHER CURVATURE TERMS}
The model we propose composes from Einstein's gravity in the presence of a scalar field and two types of higher curvature corrections, Gauss-Bonnet and Chern-Simons terms. The gravitational action is,
\begin{equation}
\centering
\label{action}
S=\int {d^4x\sqrt{-g}\left( \frac{h(\phi)R}{2\kappa^2}-\frac{1}{2} g^{\mu\nu}\partial_\mu\phi\partial_\nu\phi-V(\phi)-\xi(\phi)\mathcal{G}+\frac{1}{8}\nu(\phi)R\Tilde{R}\right)},
\end{equation}
where $h(\phi)$ is a dimensionless scalar function coupled to the Ricci scalar, $g$ is the determinant of the metric tensor, $\kappa=\frac{1}{M_P}$ is the gravitational constant while, $M_P$ denotes the reduced Planck mass, $V(\phi)$ is the scalar potential, $\xi(\phi)$
signifies the Gauss-Bonnet coupling scalar function and $\nu(\phi)$ is the Chern-Simons coupling scalar function. The Chern-Simons term represents parity violation in gravity and it is given by the expression  $R\Tilde{R}=\epsilon^{a bcd}R^{\ \ e f}_{a b}R_{cdef}$ where, $\epsilon^{abcd}$ is the totally antisymmetric Levi-Civita tensor in 4-dimensions. The Gauss-Bonnet term $\mathcal{G}$ involves a combination of quadratic curvature terms and it is given by the expression $\mathcal{G}=R^2-4R_{\alpha\beta}R^{\alpha\beta}+R_{\alpha\beta\gamma\delta}R^{\alpha\beta\gamma\delta}$,
with $R_{\alpha\beta}$ and $R_{\alpha\beta\gamma\delta}$ being the
Ricci and Riemann tensor respectively. Furthermore, the line-element is assumed to have the
Friedmann-Robertson-Walker form,
\begin{equation}
\centering
\label{metric}
ds^2=-dt^2+a(t)^2\delta_{ij}dx^idx^j,\,
\end{equation}
where $a(t)$ is the scale factor of the Universe and the metric tensor has the form of $g_{\mu\nu}=diag(-1, a(t)^2, a(t)^2, a(t)^2)$.  As long as the metric is flat, the Ricci scalar and the Gauss-Bonnet term are topological invariant and can be written as $R=12H^2+6\dot H$,
$\mathcal{G}=24H^2(\dot H+H^2)$ respectively. $H$ is Hubble's parameter and in addition, the ``dot'' denotes differentiation with respect to the cosmic time. The equations of motion for the theory can be derived by implementing the variational principle in the gravitational action (\ref{action}) with respect to the metric tensor and the scalar field separately which reads,
\begin{equation}
\centering
\label{motion1}
\frac{3hH^2}{\kappa^2}=\frac{1}{2}\dot\phi^2+V-\frac{3H\dot h}{\kappa^2}+24\dot\xi H^3,\,
\end{equation}
\begin{equation}
\centering
\label{motion2}
-\frac{2h\dot H}{\kappa^2}=\dot\phi^2-\frac{H\dot h}{\kappa^2}-16\dot\xi H\dot H+\frac{h''\dot \phi^2+h' \ddot \phi}{\kappa^2}-8H^2(\ddot\xi-H\dot\xi),\,
\end{equation}
\begin{equation}
\centering
\label{motion3}
\ddot\phi+3H\dot\phi+V'-\frac{Rh'}{2\kappa^2}+24\xi' H^2(\dot H+H^2)=0.\
\end{equation}
  The first being the Friedmann equation, the second being the Raychadhuri equation and the third is the continuity equation for the scalar field or in other words the Klein-Gordon equation for an expanding background. These equation have already been extracted in \cite{Hwang:2005hb} by Hwang and Noh in detail. The “prime” denotes differentiation with respect to the scalar field $\phi$. For the case of homogeneous and isotropic metric the Chern-Simons term and specifically the scalar coupling function $\nu(\phi)$ does not affect the background equations directly, however its impact is reflected on the behavior of the tensor perturbations and in particular in the tensor-to-scalar ratio and the tensor spectral index, which is an interesting and rather convenient characteristic of this term. Solving this particular system of differential
equations requires finding an analytical expression for Hubble’s parameter and the scalar field $\phi$, which should give us a complete description of the inflationary era.  Unfortunately, these equations are very complicated and the system
cannot be solved analytically. The solution may be extracted, only if certain approximations are made which facilitate
our study and in fact make the system solvable. Before we proceed with the approximations, we shall impose a strong
constraint on the velocity of the gravitational waves in order to achieve compatibility with the recent GW170817
observation.

Gravitational waves are perturbations in the metric which travel through spacetime with the speed of light.
Recently, in our previous work \cite{Odintsov:2020sqy} proved that compatibility with recent striking observations from GW170817 can be achieved after implementing certain additional conditions in the speed of the gravitational wave. According to Hwang and Noh paper \cite{Hwang:2005hb}, the gravitational wave speed in
natural units for Einstein-Gauss-Bonnet theories has the form,
\begin{equation}
\centering
\label{cTeq}
c_T^2=1-\frac{Q_f}{2Q_{t_{GB}}},\,
\end{equation}
where $Q_f= 16(\ddot\xi-H\dot\xi)$ and $Q_{t_{GB}}$, to be specified in the following, are auxiliary functions. If gravitons are massless during and after the inflationary era, compatibility can be achieved by equating the velocity of gravitational waves with unity, or making it infinitesimally close to unity. In other words, we demand $Q_f=0$. We select to proceed with the nontrivial case of $\xi(\phi)$ being finite instead of 0 in order to include such string corrections in subsequent results. Although it turns out that the numerical value of such function is quite negligible during the first horizon crossing, the important ratio $\frac{\xi'}{\xi''}$ as we shall indicate subsequently affects essentially the evolution of the scalar field and thus the observed indices are in an essence influenced by such scalar function. The constraint leads to an ordinary differential equation $\ddot \xi=H\dot \xi$. Now we shall solve this equation in terms of the derivatives of scalar field. Assuming that $\dot \xi=\xi' \dot \phi$ and $\frac{d}{dt}=\dot \phi \frac{d}{d\phi}$ the constraint equation has the form,
\begin{equation}
\label{constraint}
\centering
\xi'' \dot \phi^2+\xi' \ddot \phi=H\xi' \dot \phi.
\end{equation}
Considering the slow-roll approximation during inflationary era
\begin{equation}
\centering
 \ddot \phi\ll H\dot \phi,
\end{equation}
Eq. (\ref{constraint}) can be solved easily with respect to the derivative of the scalar field,
\begin{equation}
\label{fdot}
\centering
\dot \phi \simeq \frac{H\xi'}{\xi''}.
\end{equation}
In order to study the inflationary era of the Universe it is necessary to solve analytically the system of equations of motion. It is obvious that, this system is very difficult to study analytically. Thus, we assume the slow-roll approximations during inflation. Mathematically speaking, the following conditions are assumed to hold true,
\begin{align}
\label{approx}
\centering
\dot H&\ll H^2,& \frac{1}{2}\dot\phi^2&\ll V,& \ddot\phi\ll3 H\dot\phi,\
\end{align}
thus, the equations of motion can be simplified greatly. 
Hence, after imposing the constraint of the gravitational wave and considering the slow-roll approximations, the equations of motion have the following elegant forms,
\begin{equation}
\centering
\label{motion1a}
\frac{3hH^2}{\kappa^2}\simeq V-\frac{3H h' \dot \phi }{\kappa^2}+24\dot\xi H^3,\,
\end{equation}
\begin{equation}
\centering
\label{motion2a}
-\frac{2h\dot H}{\kappa^2}\simeq \dot\phi^2-\frac{H h' \dot \phi}{\kappa^2}-16\dot\xi H\dot H+\frac{h''\dot \phi^2}{\kappa^2},\,
\end{equation}
\begin{equation}
\centering
\label{motion3a}
3H\dot\phi+V'-\frac{6 H^2 h'}{\kappa^2}+24\xi' H^4\simeq0,\
\end{equation}
However, even with the slow-roll approximations holding true, the system of differential equations still remains intricate
and cannot be solved. Further approximations are needed in order to derive the inflationary phenomenology, so  we neglect string corrections themselves. This is a reasonable assumption since even though the Gauss-Bonnet scalar coupling function is seemingly neglected, it participates indirectly from the gravitational wave condition. Also, in many cases, string corrections are proven to be subleading. In addition, by using the Eq. (9) for the derivative of the
scalar field the equations of motion are simplified as,
\begin{equation}
\centering
\label{motion1b}
\frac{3hH^2}{\kappa^2}\simeq V-\frac{3H^2h' }{\kappa^2}\frac{\xi '}{\xi ''},\,
\end{equation}
\begin{equation}
\centering
\label{motion2b}
-\frac{2h\dot H}{\kappa^2}\simeq H^2\left(\frac{\xi '}{\xi''}\right)^{2} -\frac{H^2 h'} {\kappa^2}\frac{\xi '}{\xi ''}+\frac{h''H^2}{\kappa^2}\left(\frac{\xi '}{\xi ''}\right)^2,\,
\end{equation}
\begin{equation}
\centering
\label{motion3b}
3H^2\frac{\xi '}{\xi ''}+V'-\frac{6H^2 h'}{\kappa^2}\simeq0\
\end{equation}

The dynamics of inflation can be described by six parameters named
the slow-roll indices, defined as follows
\cite{Hwang:2005hb,Odintsov:2020sqy},
\begin{align}
\centering \epsilon_1&=-\frac{\dot
H}{H^2},&\epsilon_2&=\frac{\ddot\phi}{H\dot\phi},&\epsilon_3&=\frac{\dot
F}{2HF},&\epsilon_4&=\frac{\dot E}{2HE},&\epsilon_5&=\frac{\dot
F+Q_a}{2HQ_{tGB}},&\epsilon_6&=\sum_{L,R}\frac{\dot Q_t}{2HQ_t},
\end{align}
where summation over $L$ and $R$ implies summation over left and right handed polarization of gravitational waves. Parameter $Q_t$ is derived from the presence of the two curvature corrections in the mode equation for tensor perturbations defined as,
\begin{equation}
 Q_t=Q_{t{_{GB}}}+Q_{t{_{CS}}},\ 
\end{equation}
specifically $Q_{t_{GB}}=\frac{F}{\kappa^2}-8\dot\xi H$ stands for the contribution of the Gauss-Bonnet term, $Q_{t_{CS}}=\frac{F}{\kappa^2}+2\frac{\lambda_l\dot v k}{a}$ is the Chern-Simons contribution and finally $F=\frac{h(\phi)}{\kappa^2}$. Such auxiliary functions are important for studying the behavior of the gravitational wave modes. The analysis has already been performed in \cite{Hwang:2005hb}. Overall, the mode equation reads

\begin{equation}
    \centering
    \label{modeeq}
    \frac{1}{a^3Q_t}\frac{d}{dt}\left(a^3Q_t\dot h_{l \vec{k}}\right)+c_T^2\frac{k^2}{a^2}h_{l \vec{k}}\, ,
\end{equation}
with $h_{l \vec{k}}$ being the amplitude of a $\vec{k}$ mode with $l$ polarization. The above equation suggests that gravitational waves are essentially affected by the inclusion of two string corrective terms. First of all, the Gauss-Bonnet term affects, as mentioned before, the velocity of gravitational waves which can be set equal to unity in natural units following a nontrivial solution of the differential equation (\ref{cTeq}) and secondly the Chern-Simons term which indicates different behavior of a k mode depending on the polarization. In consequence, the tensor modes are strongly affected which implies different dependence of the tensor-to-scalar and the tensor spectral index of primordial curvature perturbations. Due to the small magnitude of $\xi$, as we shall indicate subsequently, the tensor modes can be approximated by their respective forms for the case of $\xi(\phi)=0$ however the finite ratio $\frac{\xi'}{\xi''}$ still affects implicitly the numerical values of such indices and in turn can result in compatible with the Planck data results. Let us now return to the slow-roll indices and the auxiliary parameters that facilitate the study.

The first two slow roll indices are used in the minimally coupled scalar field. The indices $\epsilon_3$ and $\epsilon_4$ involve the additional degree of freedom of $F(\phi)$. The index $\epsilon_5$ involves the degree of freedom of the Gauss-Bonnet coupling function $\xi(\phi)$ and finally the index $\epsilon_6$ contains the degrees of freedom of both of the coupling functions $\xi(\phi)$ and $\nu(\phi)$. The auxiliary functions are given by the following expressions,
\begin{align}
\centering Q_a&=-8\dot\xi H^2,&Q_e&=-32\dot\xi\dot H,&E&=\frac{F}{\dot \phi^2}\left(\dot \phi^2+3\frac{(\dot F +Q_a)^2}{2Q_{t_{GB}}}\right)\,
\end{align}
The first three slow-roll indices for unspecified coupling functions have quite simple forms as presented,
\begin{equation}
\centering
\label{epsilon1}
\epsilon_1=-\frac{h' \xi '}{2 h\xi ''},\
\end{equation}
\begin{equation}
\centering
\epsilon_2=1+\frac{\xi ' \left(h' \xi ''-2 h \xi ^{'''}\right)}{2 h \xi ''^2},\
\end{equation}
and $\epsilon_3=-\epsilon_1$ while, the indices $\epsilon_4$-$\epsilon_6$ have quite perplexed form due to the string-corrections involvement. Based on Ref.\cite{Hwang:2005hb}, the total auxiliary function $Q_t$ is given by the following expression,
\begin{equation}
\centering
\label{Qt}
Q_{t}=F-8\dot \xi H+2\frac{\lambda_l\dot\nu k}{\alpha},\  
\end{equation}
where the parameter $\lambda_l$ in Eq. (\ref{Qt}) represents the polarization of the primordial gravitational wave with wavenumber k and
takes values $\lambda_L=-1$ and $\lambda_R=1$ for left and right handed polarization states respectively and $\alpha$ is the scale factor.

\section{MODELS WITH HIGHER CURVATURE CORRECTIONS
COMPATIBLE WITH PLANCK DATA}
In order to ascertain the validity of a model, the results which the model produces must be confronted to the recent
Planck observational data \cite{{Akrami:2018odb}}. In the following models, we shall derive the values for the quantities, namely the
spectral index of primordial curvature perturbations $n_\mathcal{S}$, the tensor-to-scalar-ratio r and finally, the tensor spectral
index $n_\mathcal{T}$. These quantities are connected with the slow-roll indices introduced previously, as shown below,
\begin{align}
\label{observed}
\centering
n_s&=1-2\frac{2\epsilon_1+\epsilon_2-\epsilon_3+\epsilon_4}{1-\epsilon_1},&n_T&=-2\frac{\epsilon_1+\epsilon_6}{1-\epsilon_1},&r&=16\left|\left(\frac{\epsilon_1+\epsilon_3}{2}\sum_{\lambda=L,R}\left|\frac{Q_{tGB}}{Q_t}\right|-\frac{Q_e}{4HF}\right)\frac{F c_A^3}{Q_{{tGB}}}\right|,
\end{align}
where the $c_A$ the sound wave velocity defined as,
\begin{equation}
\centering
\label{soundwave}
c_A^2=1+\frac{(\dot F+Q_a)Q_e}{2Q_{t_{GB}}\dot \phi^2+3(\dot F+Q_a)^2}.\,
\end{equation}
However, by capitalizing on the fact that the GW170817 constraint generates essentially infinitesimal string corrections to the point where only the ratio $\frac{\xi'}{\xi''}$ becomes interesting, the form of the tensor-to-scalar ratio can be simplified to a great extend. Assuming that for each $\xi(\phi)$ term the contribution is quite negligible, then heuristically speaking $F\simeq Q_{tGB}$. Subsequently the field propagation velocity becomes $c_A\simeq1$ and thus one can use the tensor-to-scalar ratio for a pure Chern-Simons term that would appear for a simple canonical scalar field, which reads

\begin{equation}
    \centering
    \label{rapprox}
    r=8\left|\epsilon_1+\epsilon_3\right|\sum_{L,R}\left|\frac{F}{Q_{tCS}}\right|\, ,
\end{equation}
since $Q_t\simeq Q_{tCS}$ for weak $\xi(\phi)$. In subsequent we make comparison between the 2 forms however by ascertaining the validity of the slow-roll assumptions at the end of each seperate model it becomes abundantly clear that indeed the 2 forms essentially coincide. As a final note, it is necessary to argue that since we wish to extract results during the first horizon crossing, one has the liberty of replacing the ratio $\frac{k}{a}$ with Hubble's parameter in $Q_{tCS}$.

Based on the latest Planck observational data \cite{{Akrami:2018odb}} the spectral index of primordial curvature  perturbations is $n_\mathcal{S}=0.9649\pm 0.0042 $
and the tensor-to-scalar-ratio r must be $r<0.064$.
Our goal now is to evaluate the observational
indices during the first horizon crossing. However, instead of
using wavenumbers, we shall use the values of the scalar potential
during the initial stage of inflation. Taking it as an input, we
can obtain the actual values of the observational quantities. We
can do so by firstly evaluating the final value of the scalar
field. This value can be derived by equating slow-roll index
$\epsilon_1$ in equation (\ref{epsilon1}) to unity. Consequently, the
initial value can be evaluated from the $e$-folding number,
defined as
$N=\int_{t_i}^{t_f}{Hdt}=\int_{\phi_i}^{\phi_f}{\frac{H}{\dot\phi}d\phi}$,
where the difference $t_f-t_i$ signifies the duration of the
inflationary era. Recalling the definition of $\dot\phi$ in Eq.
(\ref{fdot}), one finds that the proper relation from which the
initial value of the scalar field can be derived is,
\begin{equation}
\centering
\label{efolds1}
N=\int_{\phi_i}^{\phi_f}{\frac{\xi''}{\xi'}d\phi}.\,
\end{equation}
From this equation, as well as equation (\ref{epsilon1}), it is
obvious that choosing an appropriate coupling function $\xi(\phi)$, is the key
in order to simplify the results.

Before we proceed with the presentation of some the results for some simple models of interest, let us return briefly to Eq. (\ref{soundwave}). As showcased, the field propagation velocity captures scalar perturbations and therefore is not affected by the Chern-Simons scalar coupling function $\nu(\phi)$. Furthermore, recall that having a nontrivial $\xi(\phi)$ solution from the differential equation $\ddot\xi=H\dot\xi$ extracted by relying on the slow-roll assumption of the scalar field suggests a really small value of $\xi(\phi)$ during the first horizon crossing, something which will become apparent in the following models. As a result, the aforementioned velocity receives contribution mainly from the non-minimal function $h(\phi)$ and therefore under the slow-roll assumption becomes lesser than unity and remains real. As it is presented in detail in \cite{Karydas:2021wmx} instabilities related to the squared sound-speed of scalar perturbations can be healed due to the combination of the non-minimal coupling and generalized non-minimal derivative coupling contribution. In our analysis, the subsequent models are essentially free of ghost instabilities and in addition respect causality. For the sake of consistency the numerical value of the sound wave velocity during the first horizon crossing shall be presented in the following models.

\subsection{A Model with Exponential $\nu(\phi)$}
In the following two subsections we shall consider that the dimensionless scalar function coupled to the Ricci scalar  is equal to unity, $h(\phi)=1$. Thus, the equations of motion are simplified as,
\begin{equation}
H^2\simeq \frac{\kappa^2 V}{3},
\end{equation}
\begin{equation}
\dot H \simeq - \frac{H^2}{2}\left(\frac{\kappa\xi'}{\xi''}\right)^2.
\end{equation}
As mentioned in the beginning of this paper the above equations of motion are derived from the gravitational action Eq. (\ref{action}). The action has certain unspecified functions, mainly the coupling functions $h(\phi)$, $\xi(\phi)$,
along with the scalar potential $V(\phi)$ and the function $\nu(\phi)$ coupled with the Chern-Simons term. Consequently, in order to derive the expression of Hubble’s parameter, these
functions must be determined.
In the following two models, we shall assume that the scalar
potential obeys a more simplified differential equation, which is,
\begin{equation}
\centering
\label{scalarminimal}
3H^2\frac{\xi '}{\xi ''}+V' \simeq 0 .
\end{equation}
This assumption is not necessary but it is convenient, since a more manageable potential may be derived, but we note that the assumption $24\xi' H^4\ll V'$ along with the slow-roll approximations (10), must hold true in order for the model
to be viable. These assumptions, in addition to the rest which shall make hereafter, will be validated if these hold
true at the end of each model.

In the first model we propose,
the scalar coupling function $\nu(\phi)$ and the potential of the scalar field are defined as follows,
\begin{equation}
\nu(\phi)=e^{-(\kappa \phi)},
\end{equation}
\begin{equation}
\centering
\label{p1}
V(\phi)= \frac{\phi^4}{(1+\gamma_1(\kappa \phi)^2)^2},
\end{equation}
where $\gamma_1$ is a dimensionless auxiliary parameter to be specified later. Solving Eq. (\ref{p1}) with respect to the Gauss-Bonnet coupling function we get,
\begin{equation}
 \xi(\phi)=\int ^{\kappa\phi } e^{-\frac{1}{16} x^2 \left(\gamma_1  x^2+2\right)}dx,
\end{equation}
where x is an auxiliary integration variable and the integration constant was set equal to unity for simplicity. In this case it becomes abundantly clear that due to the constraint on the velocity of gravitational waves, the number of free parameters decreases as now the Gauss-Bonnet scalar coupling function depends on the same auxiliary parameter $\gamma_1$ as the scalar potential. In turn, viability, if present, is dictated by a pair of parameters only, $\gamma_1$ and the e-folding number. It turns out that there exists a wide variety of values that produce results which are in agreement with the Planck data. 
\begin{figure}[t!]
\centering
\includegraphics[width=17pc]{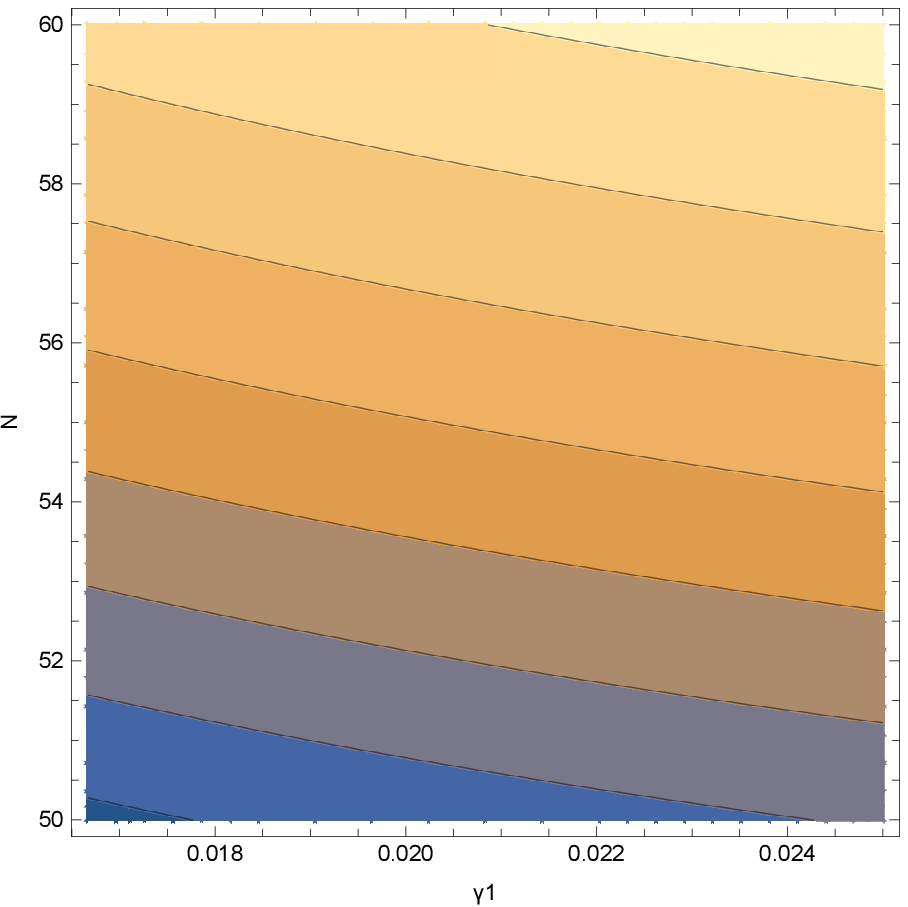}
\includegraphics[width=3pc]{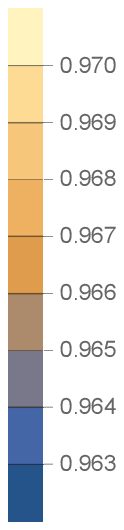}
\includegraphics[width=17pc]{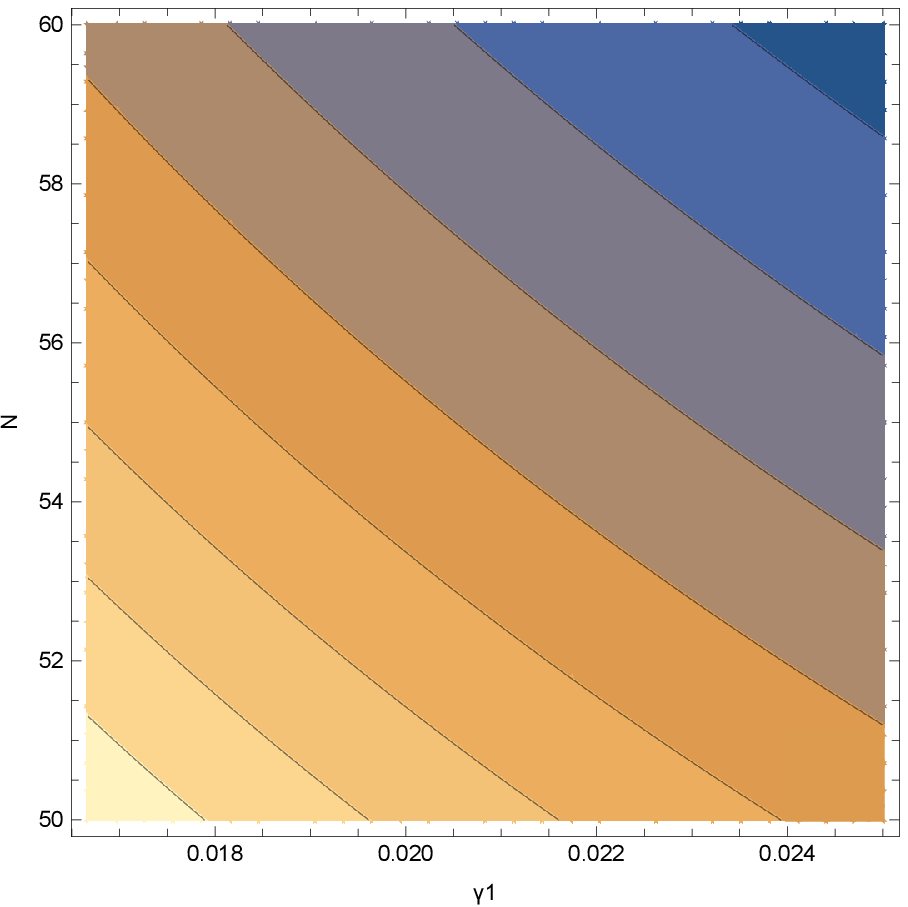}
\includegraphics[width=3pc]{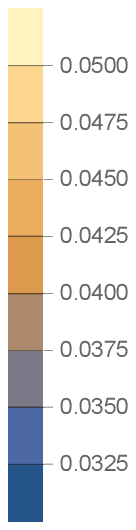}
\caption{Contour plots of the spectral index of primordial
curvature perturbations (left) and the tensor-to-scalar ratio
(right) depending on parameters $N$ and $\gamma_1$ ranging from
[50,60] and [1/60,1/40] respectively. It can be inferred that both
parameters influence their values but the spectral index changes
with a lesser rate.} \label{plot1}
\end{figure}

Concerning the slow-roll
indices, the first two are,
\begin{equation}
\epsilon_1=\frac{8}{ \left(\gamma_1 (\kappa \phi )^3+\kappa\phi \right)^2},\
\end{equation}
\begin{equation}
\epsilon_2=\frac{12 \gamma_1 \kappa ^2 \phi ^2-4}{\left(\gamma_1 (\kappa\phi)^3+\kappa  \phi \right)^2}.\
\end{equation}
As mentioned before, the first two slow-roll indices have quite simple forms, $\epsilon_3=0$ while, the rest are intricate. Let us now continue with the evaluation of the necessary values of
the inflaton field. Firstly, the final value of the scalar field can be extracted by equating index
$\epsilon_1$ to unity. As a result, the final value of the scalar field has the following form,
\begin{equation}
\phi_f=\sqrt{\frac{\left(\sqrt[3]{\gamma_1^3 (108 \gamma_1+1) \kappa ^{12}+6 \sqrt{6} \sqrt{\gamma_1^7 (54 \gamma_1+1) \kappa ^{24}}}-\gamma_1 \kappa ^4\right)^2}{3\gamma_1^2 \kappa ^6 \sqrt[3]{\gamma_1^3 (108 \gamma_1+1) \kappa ^{12}+6 \sqrt{6} \sqrt{\gamma_1^7 (54 \gamma_1+1) \kappa ^{24}}}}}.
\end{equation}
Utilizing the form of the e-folding number in Eq. (\ref{efolds1}), the initial value of the scalar field is extracted and subsequently
the observed quantities. The initial value reads,
\begin{equation}
\phi_i=\sqrt{\frac{3 \sqrt{\kappa ^4 \left(\frac{\left(\gamma_1 \kappa ^4 \sqrt[3]{\gamma_1^3 (108  \gamma_1+1) \kappa ^{12}+6 \sqrt{6} \sqrt{\gamma_1^7 (54  \gamma_1+1) \kappa ^{24}}}+\left( \gamma_1^3 (108 \gamma_1+1) \kappa ^{12}+6 \sqrt{6} \sqrt{\gamma_1^7 (54 \gamma_1+1) \kappa ^{24}}\right)^{2/3}+\gamma_1^2 \kappa ^8\right)^2}{9  \gamma_1^2 \kappa ^8 \left(\gamma_1^3 (108 \gamma_1+1) \kappa ^{12}+6 \sqrt{6} \sqrt{\gamma_1^7 (54  \gamma_1+1) \kappa ^{24}}\right)^{2/3}}+16 \gamma_1 N\right)}-3 \kappa ^2}{3\gamma_1 \kappa ^4}}.
\end{equation}
Specifying the free parameters of the theory one could produce results compatible with the observational values for the
spectral indices and the tensor-to-scalar ratio introduced in Eq. (\ref{observed}). Assuming that (N, $\gamma_1$)=(50,1/50), in reduced Planck units, so for $\kappa^2$=1, the model at hand produces acceptable results, since $n_\mathcal{S}=0.963394$ and $r=0.0469811$ are both compatible with observations. Furthermore, the tensor spectral
index takes the value $n_{\mathcal{T}}=-0.00587263$. The initial and the final numerical values of the scalar field are respectively $\phi_i=12.5609$ and $\phi_f=2.51157$ which means that the field decreasing as time flows. The numerical values of the slow-roll indices are $\epsilon_1= 0.00293632$, $\epsilon_2=0.0124302$, $\epsilon_3=0$, $\epsilon_4 \sim  \mathcal{O}(10^{-18})$, $\epsilon_5\sim \mathcal{O}(10^{-20})$ and $\epsilon_6 \sim \mathcal{O}(10^{-9})$. As expected, the indices $\epsilon_4$-$\epsilon_6$ are negligible compared to the slow-roll parameters because involves string corrections. Furthermore, either the use of Eq.(\ref{observed}) or Eq. (\ref{rapprox}) serves as a correct choice given that the GW170817 constraint $\ddot\xi=H\dot\xi$ results in infinitesimal string contributions in the tensor to scalar ratio, which appear in the form of $Q_e$, $Q_{tGB}$ and a sound wave velocity which would deviate from unity, that is $c_A<1$. In our approach , due to infinitesimal string corrections and due to the fact that the Chern-Simons coupling does not alter the field propagation velocity, we obtain $c_A=1$. This can easily be inferred from the numerical values of the string dependent slow-roll indices presented previously.
\begin{figure}[t!]
\centering
\includegraphics[width=18pc]{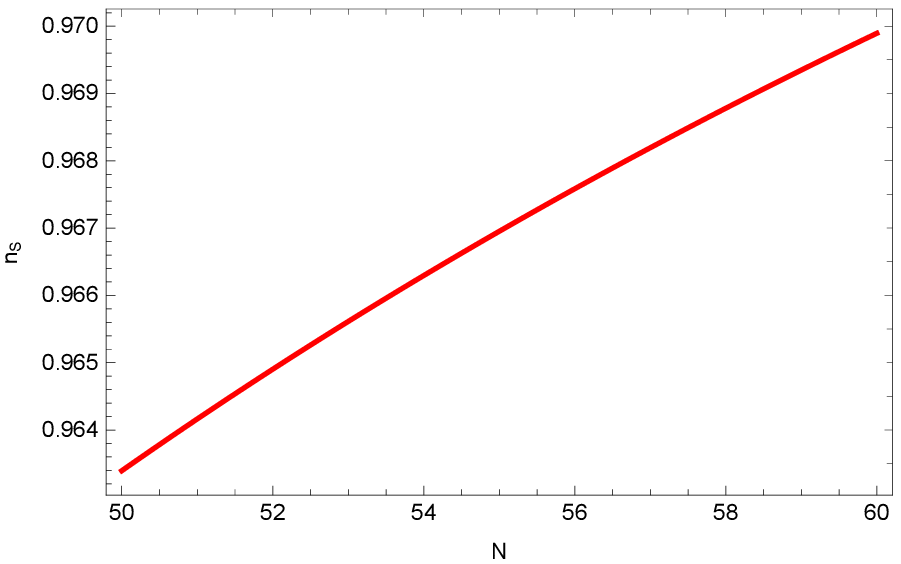}
\includegraphics[width=18pc]{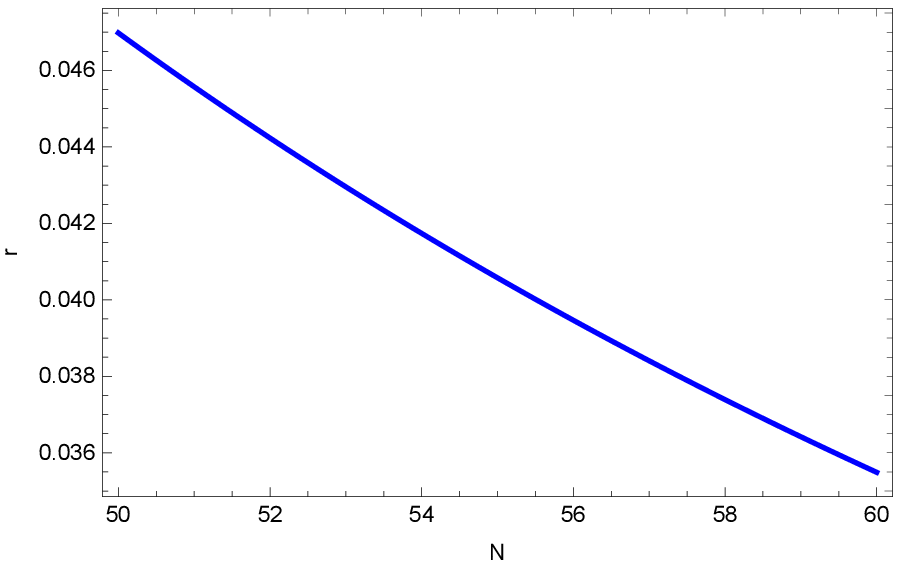}
\caption{Plots of the spectral index of primordial
curvature perturbations (left) and the tensor-to-scalar ratio
(right) depending on e-foldings number N
ranging from [50,60] for the model with exponential $\nu(\phi)$ scalar coupling function.} \label{plot2}
\end{figure}
Finally, we
examine each approximation which was made in order to derive the previous results holds true. According to the previous set of parameters in reduced Planck units always, during the first horizon crossing, $\dot H \sim \mathcal{O}(10^{1})$ and $H^2 \sim \mathcal{O}(10^{3})$  so the slow-roll assumption holds true. In addition $\frac{1}{2}\dot \phi^2 \sim \mathcal{O}(10^{1})$ while $V \sim \mathcal{O}(10^{4})$ and lastly, $\ddot \phi \sim \mathcal{O}(10^{-1})$ and $H \dot \phi \sim \mathcal{O}(10^{3})$. Hence, the slow-roll conditions are valid. All that remains is to ascertain the validity of the rest approximations. It turns out that $24\xi' H^4 \sim \mathcal{O}(10^{-16})$ which is negligible compared to $V'\sim \mathcal{O}(10^{3})$ and therefore, the
differential equation of the scalar potential is justified. Furthermore, $16\dot \xi H \dot H\sim  \mathcal{O}(10^{-20})$ is negligible compared to the term $\frac{1}{2}\dot \phi^2$ and $24\dot\xi H^3\sim\mathcal{O}(10^{-17})$ is also not significant compared to $V\sim \mathcal{O}(10^{4})$, thus the approximations in equations of motion are satisfied.

\subsection{A Model with Trigonometric potential $V(\phi)$}
Suppose now that the potential of the scalar field is defined as,
\begin{equation}
V(\phi)= sin(\gamma_2 \kappa \phi),
\end{equation}
where the amplitude of the scalar potential is assumed to be $M_P^4$, or just unity in this particular approach. The Chern-Simons scalar coupling function has the following form,
\begin{equation}
\nu(\phi)=(\kappa \phi)^{n_2}.
\end{equation}
Given the trigonometric potential of the field, the scalar Gauss-Bonnet coupling function can be derived from Eq.(\ref{scalarminimal}),
\begin{equation}
\xi(\phi)=-\frac{\gamma_2  \sqrt{\sin ^2(\gamma_2  \kappa  \phi )} \cos ^{\frac{1}{\gamma_2 ^2}+1}(\gamma_2  \kappa  \phi ) \csc( \gamma_2  \kappa  \phi ) \, _2F_1\left(\frac{1}{2},\frac{1}{2} \left(1+\frac{1}{\gamma_2 ^2}\right);\frac{1}{2} \left(3+\frac{1}{\gamma_2 ^2}\right);\cos ^2(\gamma_2  \kappa  \phi )\right)}{\gamma_2 ^2 \kappa +\kappa },\
\end{equation}
where $_2F_1\left(\frac{1}{2},\frac{1}{2} \left(1+\frac{1}{\gamma_2 ^2}\right);\frac{1}{2} \left(3+\frac{1}{\gamma_2 ^2}\right);\cos ^2(\gamma_2  \kappa  \phi )\right)$ is the hypergeometric function and the integration constant is set equal to unity such that $\gamma_2$ is the only important auxiliary parameter. At first glance the appearance of a hypergeometric function may seem intimidating however we remind the reader that Gauss-Bonnet coupling itself is not so important as only ratios $\frac{\xi'}{\xi''}$ participate usually in the equations above which can be simplified greatly. Moreover a proper designation of $\gamma_2$ in the appearance of complex numbers is avoided. Similar to the previous model, the Gauss-Bonnet scalar coupling function is again depending on the auxiliary parameter of the scalar potential so by using the constraint $\ddot\xi=H\dot\xi$ under the slow-roll assumptions decreases the effective number of parameters but in exchange a quite perplexed function emerges, at least if one assumes a periodic scalar potential. Let us now proceed with the overall phenomenology. 
The slow-roll parameters of the model are given by the following expressions,
\begin{equation}
\epsilon_1=\frac{1}{2} \gamma_2^2 \cot ^2(\gamma_2 \kappa  \phi ) ,\
\end{equation}
\begin{equation}
\epsilon_2=\frac{1}{2} \gamma_2^2 \left(\csc ^2(\gamma_2 \kappa  \phi )+1\right),\
\end{equation}
$\epsilon_3=0$ and the indices $\epsilon_4$ and $\epsilon_6$ where omitted due to their complicated form.
Similar to the previous model, the value of the scalar field at the end of inflation is derived from the equation
$\epsilon_1=1$ and therefore it reads,
\begin{equation}
\phi_f=\frac{2 \tan ^{-1}\left(\sqrt{\frac{\gamma_2^2-2 \sqrt{2} \sqrt{\gamma_2^2+2}+4}{\gamma_2^2}}\right)}{\gamma_2 \kappa}.
\end{equation}
Considering the form of the e-folding number in Eq.(\ref{efolds1}), the initial value of the scalar field is extracted and subsequently
the observed quantities. The initial value reads,
\begin{equation}
\phi_i=\frac{sec^{-1}(e^{\gamma_2^2 N}sec(\phi_f))}{\gamma_2 \kappa}.
\end{equation}
\begin{figure}[t!]
\centering
\includegraphics[width=17pc]{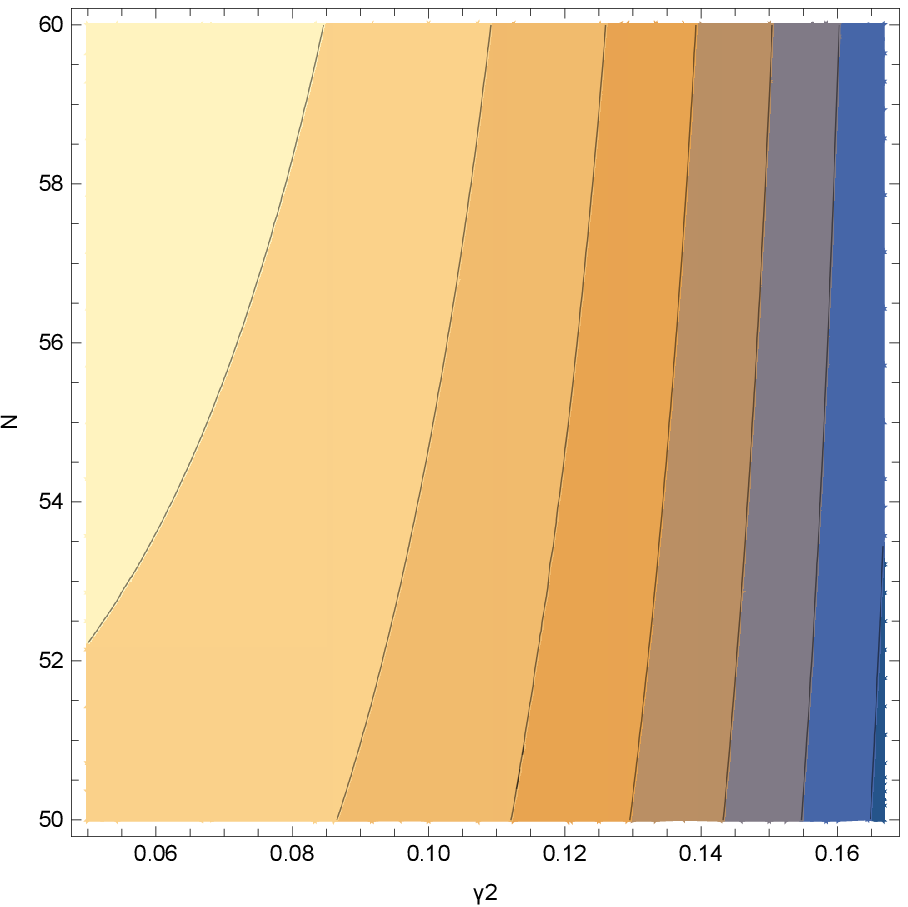}
\includegraphics[width=3pc]{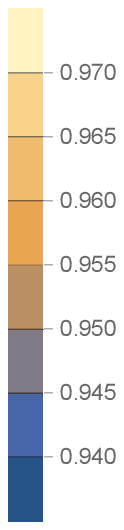}
\includegraphics[width=17pc]{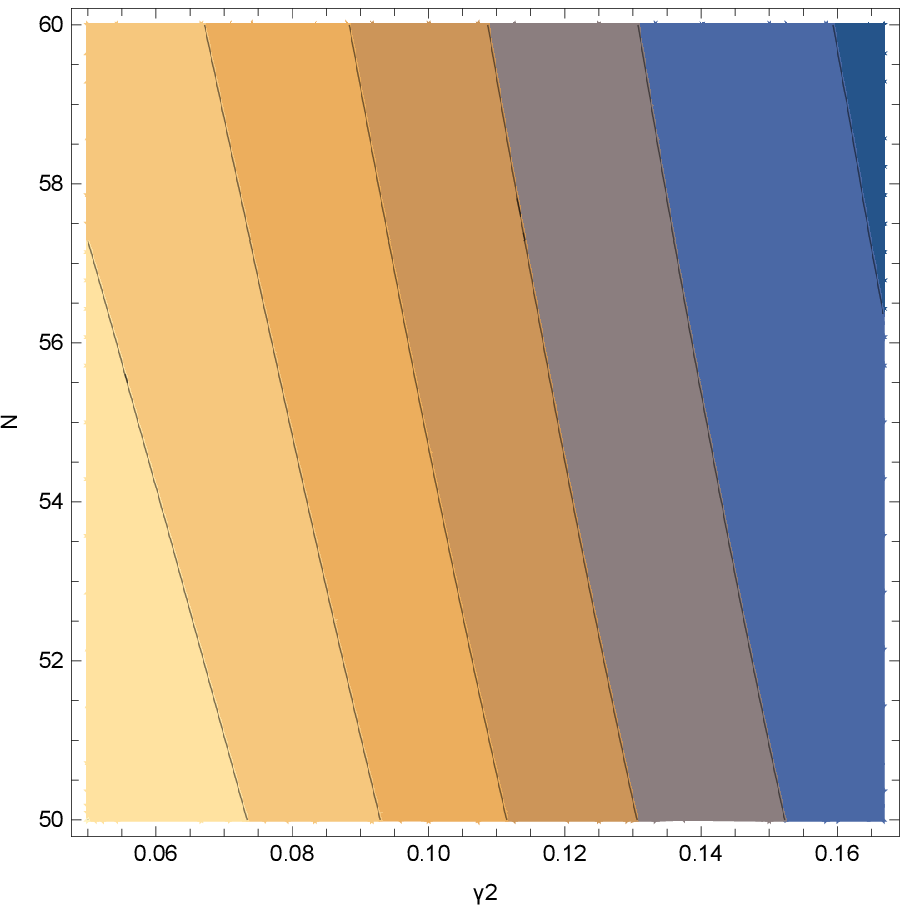}
\includegraphics[width=3pc]{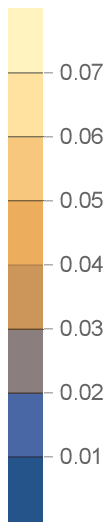}
\caption{Contour plots of the spectral index of primordial
curvature perturbations (left) and the tensor-to-scalar ratio
(right) depending on parameters $N$ and $\gamma_2$ ranging from
[50,60] and [1/20,1/6] respectively. } \label{plot2}
\end{figure}
Assuming that in reduced Planck Units, the free parameters have the following values (N, $\gamma_2$, $n_2$)=(60, 1/9, 1) we obtain viable results for the observational quantities, which are in good agreement
with experimental evidence \cite{Akrami:2018odb}, since $n_\mathcal{S}=0.9645$ and $r=0.0288588$. Hence, the trivial case of a linear Chern-Simons coupling suffices in order to obtain compatible with the observations results. In addition, the tensor spectral index is $n_{\mathcal{T}}=-0.00357223$ and $c_A=1$ which means that the model is free of ghosts.  Concerning the scalar field itself, we mention that in Planck Units, $\phi_i=9.67946$ and $\phi_f=0.705657$  which indicates
a decreasing with time homogeneous scalar field. Moreover, for the numerical values of the slow-roll indices, we
mention that $\epsilon_1=0.00180144$, $\epsilon_2=0.0141471$, $\epsilon_3=0$, $\epsilon_4 \sim  \mathcal{O}(10^{-16})$, $\epsilon_5\sim \mathcal{O}(10^{-28})$ and $\epsilon_6 \simeq -0.000015322$. Referring to the choice of the tensor-to-scalar ratio, it is worth mentioning that indeed no matter the choice of the form suggested in the beginning of the section the results are indistinguishable and thus using the one for a pure Chern-Simons contribution for a canonical scalar field is indeed a viable option. This statement is also supported from the numerical values of the pure string corrective terms which shall be presented shortly. Also, parameter $n_2$ as expected does not influence the scalar spectral index since it participates in the tensor perturbations only, as mentioned previously. 
\begin{figure}[t!]
\centering
\includegraphics[width=20pc]{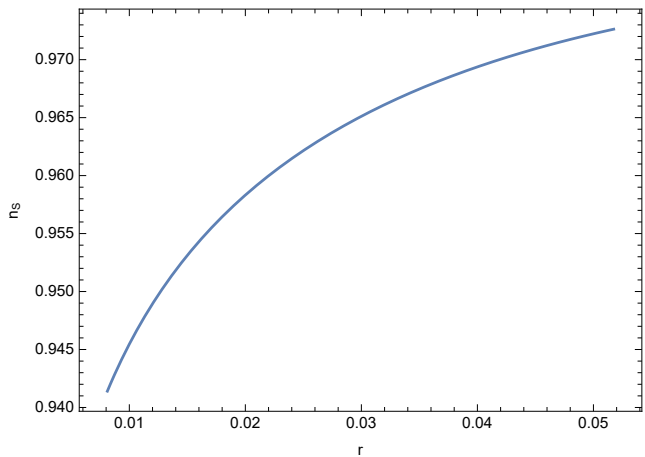}
\caption{Parametric plot of the spectral index of primordial
curvature perturbations as function of tensor-to-scalar ratio
 for the model with trigonometric potential for $N=60$ e-foldings. The parameter $\gamma_2$ for the indices is ranging from [1/20,1/6]. } \label{plot4}
\end{figure}

Without a doubt, the dominant contribution in both the tensor-to-scalar ratio and the slow-roll index $\epsilon_6$ comes from the Chern-Simons term. Essentially one could argue that the apparent dominance of the Chern-Simons term relative to string corrective term proportional to $\xi(\phi)$ and not the ratio $\frac{\xi'}{\xi''}$ could generate a blue shifted tensor spectral index. Indeed this is a plausible scenario that is further investigated in the third model where a relatively trivial non-minimal coupling between the Ricci scalar and the scalar field is assumed.

Finally, we examine the validity of the approximations which were made necessarily in order to solve approximately
the system of equations of motion. Referring to the slow-roll approximations which were assumed, we mention that
during the first horizon crossing, $\dot H \sim \mathcal{O}(10^{-4})$ and $H^2 \sim \mathcal{O}(10^{-1})$,  so the condition $\dot H<<H^2$ holds true. Furthermore
 $\frac{1}{2}\dot \phi^2 \sim \mathcal{O}(10^{-4})$ while $V \sim \mathcal{O}(10^{-1})$ thus, $\frac{1}{2}\dot \phi^2\ll V$
 and finally, $\ddot \phi \sim \mathcal{O}(10^{-3})$ and $H \dot \phi \sim \mathcal{O}(10^{-2})$, it is  clear that also the condition $\ddot \phi \ll H \dot \phi$ holds true.
Hence, the slow-roll conditions are valid in this model. Now let us see whether
the rest of the assumptions made in the previous sections indeed hold true. In particular, for the first equation
of motion  we have $24\dot \xi H^3 \sim \mathcal{O}(10^{-28})$ compared to $V\sim \mathcal{O}(10^{-1})$ therefore, the condition $24 \dot \xi H^3\ll V$ holds true. Furthermore, we shall check the approximations in the second equation of motion $16\dot \xi H \dot H\sim  \mathcal{O}(10^{-30})$ is negligible compared to the term $\frac{1}{2}\dot \phi^2$ hence, the approximation $16\dot \xi H \dot H\ll \dot \phi^2$ is also valid. Lastly, in the equation of motion of the scalar field the term $24\xi'H^4\sim\mathcal{O}(10^{-26})$ is minor compared to the term $V'\sim  \mathcal{O}(10^{-2})$ thus, $24\xi'H^4\ll V'$.

\subsection{Model with a Non-Minimally scalar coupling function with Einstein's gravity}
Let us now present a model where the scalar coupling functions are defined as,
\begin{equation}
h(\phi)=\gamma_3(\kappa \phi)^m ,\
\end{equation}
\begin{equation}
\nu(\phi)=\frac{\phi}{M_P} ,\
\end{equation}
\begin{equation}
\xi(\phi)=\frac{2}{\sqrt{\pi}}\int_0^{\delta \kappa \phi} e^{-x^2}dx ,\
\end{equation}
where $\gamma_3$, $m$ and $\delta$ are dimensionless parameters while x serves as an auxiliary integration variable.
The choice of an error function is known for describing a viable model and subsequently results in a simple ratio between the first two derivatives of the Gauss-Bonnet scalar coupling function. As a result, the scalar potential is expected to obtain a simple exponential like form. The equations of motion are simplified as,
\begin{equation}
H^2 \simeq \frac{\kappa^2 V}{3h},\
\end{equation}
\begin{equation}
\label{dotHapprox}
\dot H \simeq \frac{H^2}{2}\frac{h'}{h}\frac{\xi '}{\xi ''},\
\end{equation}
\begin{equation}
\centering
\label{potentialnon}
V'+3H^2\frac{\xi'}{\xi''}-\frac{6H^2h'}{\kappa^2}\simeq 0.
\end{equation}
Solving Eq.(\ref{potentialnon}) with respect to  the scalar potential,
\begin{equation}
V(\phi)=V_3 \phi^{2m}\exp \left(-\frac{(\kappa  \phi )^{1-m}}{\gamma_3  (1-m)}\right),\
\end{equation}
where $V_3$ is the constant integration with mass dimensions $[m]^{4-2m}$. It is mainly introduced for dimensional purposes however hereafter we assume it is equal to unity and proceed. Hence, the trivial choices of power-law and error function scalar couplings resulted in a combination of a power-law and an exponential scalar potential. The first three slow-roll indices of the model have quite simple and elegant expressions as shown,
\begin{equation}
\epsilon_1=\frac{m}{4 \delta ^2 \kappa ^2 \phi ^2} ,\   
\end{equation}
\begin{equation}
\epsilon_2=-\frac{m-2}{4 \delta ^2 \kappa ^2 \phi ^2},\
\end{equation}
\begin{equation}
\epsilon_3=-\frac{m}{4 \delta ^2 \kappa ^2 \phi ^2},\
\end{equation}
while the rest indices are omitted due to their perplexed form.  Equating to unity the
first slow-roll index however, one obtains the following values for the scalar field,
\begin{equation}
\phi_f=-\frac{\sqrt{m}}{2 \delta  \kappa },\
\end{equation}
\begin{equation}
\phi_i=-\frac{\sqrt{\frac{m}{4}+N}}{\delta \kappa }.
\end{equation}
Obviously the inclusion of a non-minimal coupling between the  Ricci scalar and the scalar field raises the number of the free parameters but yet again, given that the aforementioned gravitational wave constraint applies one can observe dependence on the same free parameters in several scalar functions. Moreover on all of them are important, in fact compatibility can be achieved relatively easy without the need of fine tuning. In fact, by assigning the following values
to the free parameters, always in reduced Planck units ( $N$, $\delta$, $\gamma_3$, $m$)=( 60, 1, 1, 2) the observational indices take the values $n_{\mathcal{S}}=0.966942$, $n_{\mathcal{T}}=0.0165494$ and $r=0.00164156$ which are compatible with the latest observations. It is worth mentioning that designating $m=2$ suggests that the scalar potential is quite close to a $\phi^4$ potential. Truthfully speaking, the exponential part of the potential acts as a correction to the $\phi^4$ part by suppressing it as during the first horizon crossing for instance, $\phi_i^4\simeq3660$ whereas $V\simeq3218$. The scalar field seems to increase with time since $\phi_i=-7.77817$ and $\phi_f=-0.707107$. Also, the model is free of ghosts since $c_A=1$. Furthermore, the slow-roll indices have the following numerical values
$\epsilon_1=0.00826446$, $\epsilon_2 \sim \mathcal{O}(10^{-16})$, $\epsilon_3\simeq\epsilon_4\simeq\epsilon_5\simeq -0.00826446$ while $\epsilon_6 \simeq -0.0165392$.
\begin{figure}[t!]
\centering
\includegraphics[width=17pc]{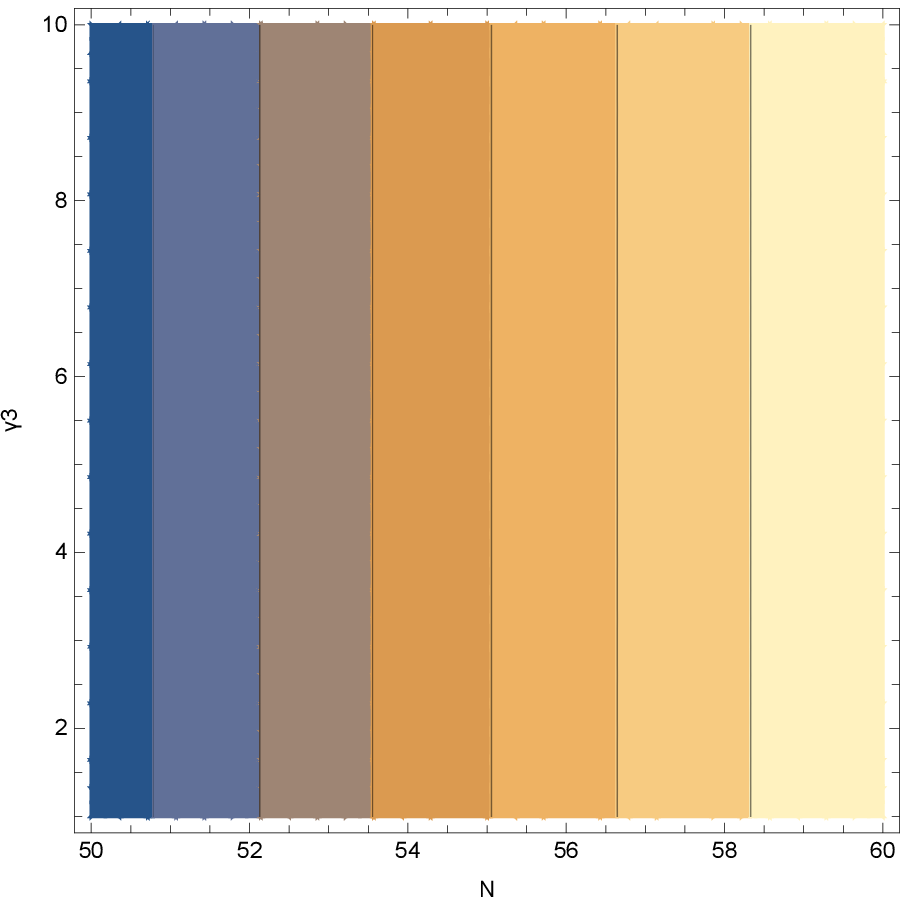}
\includegraphics[width=3pc]{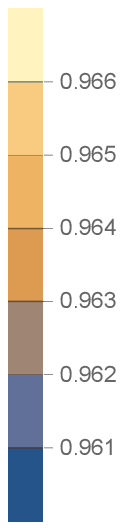}
\includegraphics[width=17pc]{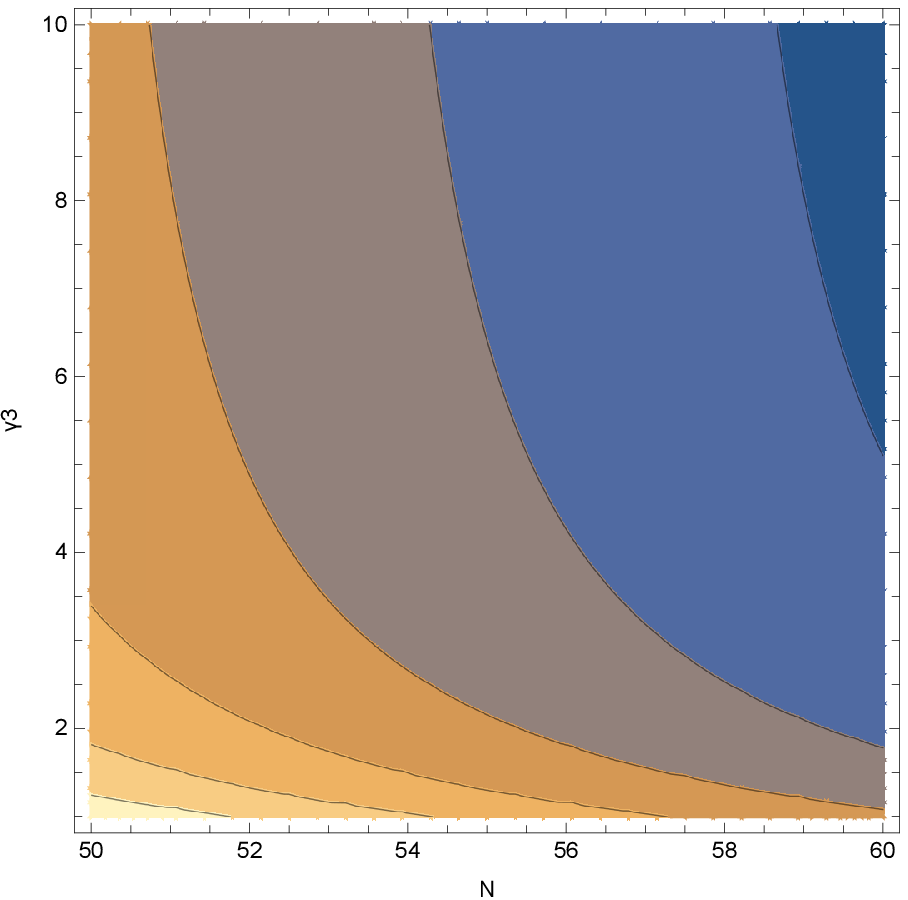}
\includegraphics[width=3pc]{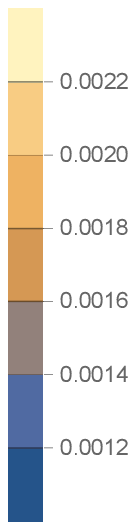}
\caption{Contour plots of the spectral index of primordial
curvature perturbations (left) and the tensor-to-scalar ratio
(right) depending on parameters $N$ and $\gamma_3$ ranging from
[50,60] and [1,10] respectively. } \label{plot3}
\end{figure}
The observant reader may realise that the value of the tensor-to-scalar ratio does not match with the numerical values of the slow-roll index, given that $\epsilon_1$ is opposite to $\epsilon_3$ thus stating that a zero tensor-to-scalar ratio is the correct result. Indeed, that would be the case if the approximations imposed on the form of $\dot H$ in Eq. (\ref{dotHapprox}) were not approximations but rather accurate results. Also, from another point of view it would be pointless to expect zero B modes in the CMB but extract a positive value of the tensor spectral index. The result for $r$ is actually finite because the next leading order in $\dot H$ was used in order to reformulate the first slow-roll index $\epsilon_1$ and thus extract a finite value. Adding only the second derivative of the Ricci coupling $h(\phi)$ which was previously dismissed suffices. In fact contribution from the kinetic term of the scalar field affects mildly insignificant decimals. The same applies obviously to further corrections in the tensor spectral index as the leading order from Eq.(\ref{dotHapprox}) is dominant. Such statement is in agreement with not only the slow-roll assumption but also with the initial choice of Hubble's time derivative.

It is worth stating that the aforementioned designation results in a blue tilted tensor spectral index. Truthfully, this is an interesting result that is generated due to the existence of the Chern-Simons term in the gravitational action (\ref{action}). Under the slow-roll assumption and for a Gauss-Bonnet term $\xi(\phi)\mathcal{G}$, compatibility with the GW170817 event suggests that pure string corrective terms are negligible compared to the Planck scale therefore the usual condition $r\simeq-8n_\mathcal{T}$ is indeed valid. Here however, it can easily be inferred that such relation is violated not only due to the sign but also from the order of magnitude as the tensor spectral index is roughly speaking one order of magnitude greater than the tensor-to-scalar ratio. This intriguing result is a direct consequence of the model and the specific choice of auxiliary parameters. From a certain point of view one could argue that such result was fine tuned. There does not exist a universal master relation that if satisfied guarantees that the tensor spectral index shall be blue tilted, other than the relation $\epsilon_6<\epsilon_1<0$. Once again, such condition is only feasible due to the Chern-Simons term.

Finally, let us discuss here and validate whether the approximations assumed in the section II hold true. First of all, we shall check the validity of the slow-roll approximations, $\dot H \sim \mathcal{O}(10^{-1})$ and $H^2\sim \mathcal{O}(10^{2})$ hence the approximation, $\dot H\ll H^2$ holds true. In addition the kinetic term of the scalar field is $\frac{1}{2}\dot \phi^2 \sim \mathcal{O}(10^{-2})$ while the scalar potential is  $V\sim\mathcal{O}(10^{4})$, hence the condition $\frac{1}{2}\dot \phi^2\ll V$ is valid. Lastly, comparing the term $\ddot \phi\sim \mathcal{O}(10^{-3})$ with the term $H \dot \phi\sim \mathcal{O}(10^{1})$, it is clear that the condition $\ddot \phi\ll 3H\dot \phi$ is satisfied. All that remains is to ascertain the validity of the rest
approximations. Concerning the first equation of motion, the terms $H \dot h \sim \mathcal{O}(10^{2})$ and $24\xi \dot H^3 \sim \mathcal{O}(10^{-24})$ are quite smaller in order of magnitude from the scalar potential V, thus our approximations holds true. In the second Eq. of motion the terms $16\dot \xi H \dot H\sim \mathcal{O}(10^{-26})$ and $h''\dot \phi^2\sim \mathcal{O}(10^{-1})$ are negligible compared to the term $H h' \dot \phi\sim \mathcal{O}(10^{2})$. About of the equation of the scalar potential the term $V'\sim \mathcal{O}(10^{4})$ is significant compared to the term
$24\xi' H^4\sim \mathcal{O}(10^{-23})$. Thus, all the approximations are valid.

\begin{figure}[t!]
\centering
\includegraphics[width=20pc]{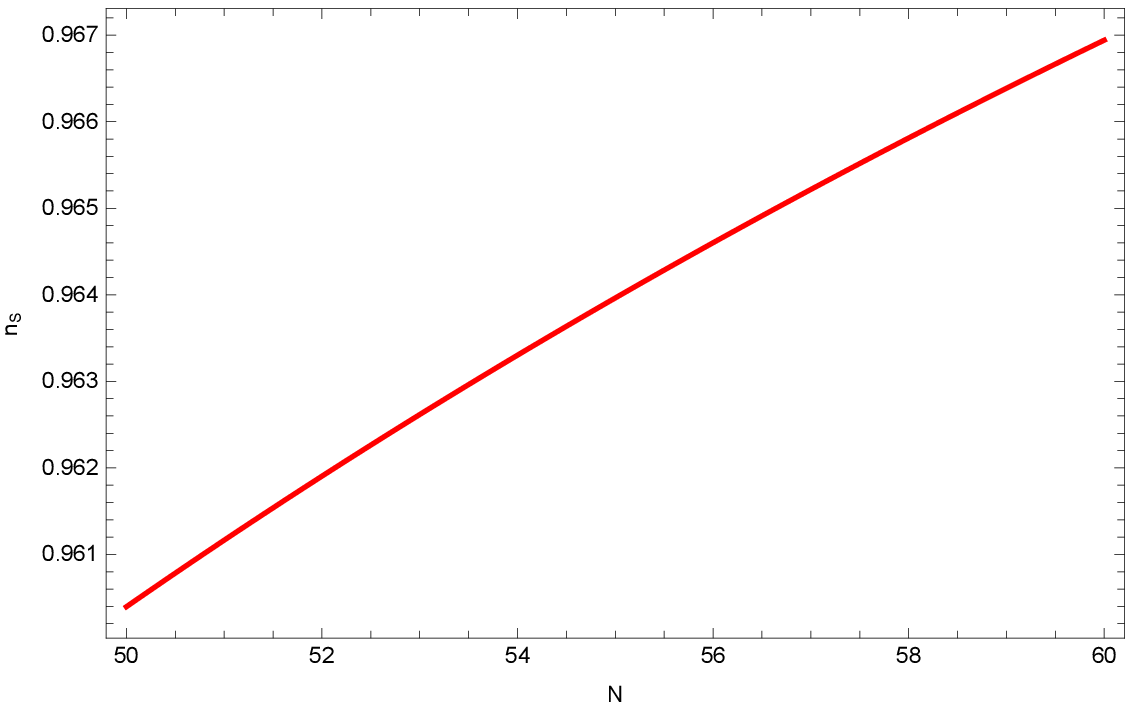}
\includegraphics[width=20pc]{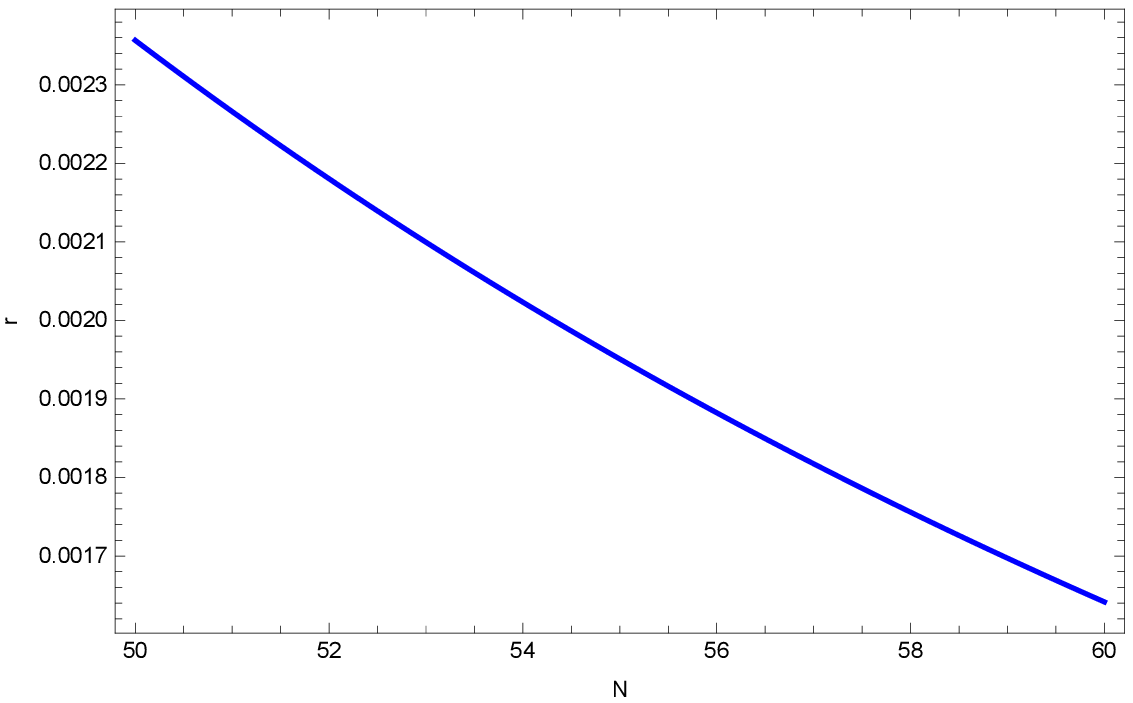}
\caption{Plots of the spectral index of primordial
curvature perturbations (left) and the tensor-to-scalar ratio
(right) depending on e-foldings number N
ranging from [50,60] for the model with a Non-Minimally scalar coupling function with Einstein’s gravity.  } \label{plot4}
\end{figure}

\section{Conclusions}
In this paper it is presented inflationary phenomenology in the context of Einstein's gravity with the existence of Gauss-Bonnet and Chern-Simons higher curvature corrections. We proved that viable models with string motivated terms can be achieved, consistent with the recent GW170817 observations  imposing the speed of the gravitational wave equal to unity.
The presence of the Chern-Simons term affected only the tensor spectral index of primordial curvature perturbations, leading to asymmetric generation and evolution of the two circular polarization states of gravitational wave, without any effect in the background equations and also in the spectral index of primordial curvature perturbations.

 Considering the constraint of the speed of the gravitational wave, quantities with different origins in the
action, became interconnected as expressed with respect to the time derivative of the scalar field. The system of differential equations, which formed from gravitational action simplified greatly, after imposing the slow-roll assumptions and ignoring string terms. The equations of motion had quite simple and elegant forms, consequently the dynamics of inflation described by the slow-roll indices using auxiliary functions.

In the last section we examined viable models consistent with observations. As demonstrated, functions which have appealing
characteristics, such as the exponential function and error function are excellent candidates for describing the inflationary era given that the ratios of the derivatives of the scalar coupling functions, which appear in the equations of motion, are greatly simplified. In the aforementioned models the expressions of the slow-roll indices and also the initial and final values of the scalar field during the inflationary era were evaluated. Our last step was to examine  the numerical values of all quantities and approximations in our models. We conclude that the models are compatible with the latest Planck data. An interesting and thus reportable result is that a blue tilted tensor spectral index is now a possibility due to the inclusion of the Chern-Simons term. Unfortunately, no master equation that guarantees such result whenever it is respected exists. The spectral index may admit a positive value however such result is model dependent and can be achieved through fine-tuning. The same applies in principle to the case of extra string corrective terms participating in the gravitational action. We leave this interesting scenario for a future work.


\begin{thebibliography}{99}
\bibitem{Nojiri:2017ncd}
S.~Nojiri, S.~D.~Odintsov and V.~K.~Oikonomou,
Phys. Rept. \textbf{692} (2017), 1-104
doi:10.1016/j.physrep.2017.06.001
[arXiv:1705.11098 [gr-qc]].

\bibitem{Capozziello:2011et}
S.~Capozziello and M.~De Laurentis,
Phys. Rept. \textbf{509} (2011), 167-321
doi:10.1016/j.physrep.2011.09.003
[arXiv:1108.6266 [gr-qc]].

\bibitem{Capozziello:2010zz}
V.~Faraoni and S.~Capozziello,
doi:10.1007/978-94-007-0165-6

\bibitem{Nojiri:2006ri}
S.~Nojiri and S.~D.~Odintsov,
eConf \textbf{C0602061} (2006), 06
doi:10.1142/S0219887807001928
[arXiv:hep-th/0601213 [hep-th]].

\bibitem{Nojiri:2010wj}
S.~Nojiri and S.~D.~Odintsov,
Phys. Rept. \textbf{505} (2011), 59-144
doi:10.1016/j.physrep.2011.04.001
[arXiv:1011.0544 [gr-qc]].

\bibitem{Olmo:2011uz}
G.~J.~Olmo,
Int. J. Mod. Phys. D \textbf{20} (2011), 413-462
doi:10.1142/S0218271811018925
[arXiv:1101.3864 [gr-qc]].



































\bibitem{Kanti:2015pda}
P.~Kanti, R.~Gannouji and N.~Dadhich,
Phys. Rev. D \textbf{92} (2015) no.4, 041302
doi:10.1103/PhysRevD.92.041302
[arXiv:1503.01579 [hep-th]].

\bibitem{Yi:2018gse}
Z.~Yi, Y.~Gong and M.~Sabir,
Phys. Rev. D \textbf{98} (2018) no.8, 083521
doi:10.1103/PhysRevD.98.083521
[arXiv:1804.09116 [gr-qc]].

\bibitem{Guo:2010jr}
Z.~K.~Guo and D.~J.~Schwarz,
Phys. Rev. D \textbf{81} (2010), 123520
doi:10.1103/PhysRevD.81.123520
[arXiv:1001.1897 [hep-th]].

\bibitem{Jiang:2013gza}
P.~X.~Jiang, J.~W.~Hu and Z.~K.~Guo,
Phys. Rev. D \textbf{88} (2013), 123508
doi:10.1103/PhysRevD.88.123508
[arXiv:1310.5579 [hep-th]].

\bibitem{Guo:2009uk}
Z.~K.~Guo and D.~J.~Schwarz,
Phys. Rev. D \textbf{80} (2009), 063523
doi:10.1103/PhysRevD.80.063523
[arXiv:0907.0427 [hep-th]].

\bibitem{DeLaurentis:2015fea}
M.~De Laurentis, M.~Paolella and S.~Capozziello,
Phys. Rev. D \textbf{91} (2015) no.8, 083531
doi:10.1103/PhysRevD.91.083531
[arXiv:1503.04659 [gr-qc]].
\bibitem{Fomin:2020hfh}
I.~Fomin,
Eur. Phys. J. C \textbf{80} (2020) no.12, 1145
doi:10.1140/epjc/s10052-020-08718-w
[arXiv:2004.08065 [gr-qc]].
\bibitem{Pozdeeva:2020apf}
E.~O.~Pozdeeva, M.~R.~Gangopadhyay, M.~Sami, A.~V.~Toporensky and S.~Y.~Vernov,
Phys. Rev. D \textbf{102} (2020) no.4, 043525
doi:10.1103/PhysRevD.102.043525
[arXiv:2006.08027 [gr-qc]].

\bibitem{Yi:2018dhl}
Z.~Yi and Y.~Gong,
Universe \textbf{5} (2019) no.9, 200
doi:10.3390/universe5090200
[arXiv:1811.01625 [gr-qc]].

\bibitem{vandeBruck:2016xvt}
C.~van de Bruck, K.~Dimopoulos and C.~Longden,
Phys. Rev. D \textbf{94} (2016) no.2, 023506
doi:10.1103/PhysRevD.94.023506
[arXiv:1605.06350 [astro-ph.CO]].

\bibitem{Odintsov:2018zhw}
S.~D.~Odintsov and V.~K.~Oikonomou,
Phys. Rev. D \textbf{98} (2018) no.4, 044039
doi:10.1103/PhysRevD.98.044039
[arXiv:1808.05045 [gr-qc]].

\bibitem{Nozari:2017rta}
K.~Nozari and N.~Rashidi,
Phys. Rev. D \textbf{95} (2017) no.12, 123518
doi:10.1103/PhysRevD.95.123518
[arXiv:1705.02617 [astro-ph.CO]].

\bibitem{Chakraborty:2018scm}
S.~Chakraborty, T.~Paul and S.~SenGupta,
Phys. Rev. D \textbf{98} (2018) no.8, 083539
doi:10.1103/PhysRevD.98.083539
[arXiv:1804.03004 [gr-qc]].

\bibitem{Kawai:1999pw}
S.~Kawai and J.~Soda,
Phys. Lett. B \textbf{460} (1999), 41-46
doi:10.1016/S0370-2693(99)00736-4
[arXiv:gr-qc/9903017 [gr-qc]].

\bibitem{vandeBruck:2017voa}
C.~van de Bruck, K.~Dimopoulos, C.~Longden and C.~Owen,
[arXiv:1707.06839 [astro-ph.CO]].





\bibitem{Bakopoulos:2020dfg}
A.~Bakopoulos, P.~Kanti and N.~Pappas,
Phys. Rev. D \textbf{101} (2020) no.8, 084059
doi:10.1103/PhysRevD.101.084059
[arXiv:2003.02473 [hep-th]].

\bibitem{Kleihaus:2019rbg}
B.~Kleihaus, J.~Kunz and P.~Kanti,
Phys. Lett. B \textbf{804} (2020), 135401
doi:10.1016/j.physletb.2020.135401
[arXiv:1910.02121 [gr-qc]].

\bibitem{Bakopoulos:2019tvc}
A.~Bakopoulos, P.~Kanti and N.~Pappas,
Phys. Rev. D \textbf{101} (2020) no.4, 044026
doi:10.1103/PhysRevD.101.044026
[arXiv:1910.14637 [hep-th]].

\bibitem{Kanti:1995vq}
P.~Kanti, N.~E.~Mavromatos, J.~Rizos, K.~Tamvakis and E.~Winstanley,
Phys. Rev. D \textbf{54} (1996), 5049-5058
doi:10.1103/PhysRevD.54.5049
[arXiv:hep-th/9511071 [hep-th]].

\bibitem{Bajardi:2019zzs}
F.~Bajardi, K.~F.~Dialektopoulos and S.~Capozziello,
Symmetry \textbf{12} (2020) no.3, 372
doi:10.3390/sym12030372
[arXiv:1911.03554 [gr-qc]].












\bibitem{Nojiri:2020pqr}
S.~Nojiri, S.~D.~Odintsov, V.~K.~Oikonomou and A.~A.~Popov,
Phys. Dark Univ. \textbf{28} (2020), 100514
doi:10.1016/j.dark.2020.100514
[arXiv:2002.10402 [gr-qc]].


\bibitem{Nojiri:2019nar}
S.~Nojiri, S.~D.~Odintsov, V.~K.~Oikonomou and A.~A.~Popov,
Phys. Rev. D \textbf{100} (2019) no.8, 084009
doi:10.1103/PhysRevD.100.084009
[arXiv:1909.01324 [gr-qc]].

\bibitem{Odintsov:2019evb}
S.~D.~Odintsov and V.~K.~Oikonomou,
Phys. Rev. D \textbf{99} (2019) no.10, 104070
doi:10.1103/PhysRevD.99.104070
[arXiv:1905.03496 [gr-qc]].




\bibitem{Odintsov:2019mlf}
S.~D.~Odintsov and V.~K.~Oikonomou,
Phys. Rev. D \textbf{99} (2019) no.6, 064049
doi:10.1103/PhysRevD.99.064049
[arXiv:1901.05363 [gr-qc]].


\bibitem{Alexander:2009tp}
S.~Alexander and N.~Yunes,
Phys. Rept. \textbf{480} (2009), 1-55
doi:10.1016/j.physrep.2009.07.002
[arXiv:0907.2562 [hep-th]].

\bibitem{Qiao:2019hkz}
J.~Qiao, T.~Zhu, W.~Zhao and A.~Wang,
Phys. Rev. D \textbf{101} (2020) no.4, 043528
doi:10.1103/PhysRevD.101.043528
[arXiv:1911.01580 [astro-ph.CO]].

\bibitem{Nishizawa:2018srh}
A.~Nishizawa and T.~Kobayashi,
Phys. Rev. D \textbf{98} (2018) no.12, 124018
doi:10.1103/PhysRevD.98.124018
[arXiv:1809.00815 [gr-qc]].

\bibitem{Wagle:2018tyk}
P.~Wagle, N.~Yunes, D.~Garfinkle and L.~Bieri,
Class. Quant. Grav. \textbf{36} (2019) no.11, 115004
doi:10.1088/1361-6382/ab0eed
[arXiv:1812.05646 [gr-qc]].


\bibitem{Yagi:2012vf}
K.~Yagi, N.~Yunes and T.~Tanaka,
Phys. Rev. Lett. \textbf{109} (2012), 251105
[erratum: Phys. Rev. Lett. \textbf{116} (2016) no.16, 169902; erratum: Phys. Rev. Lett. \textbf{124} (2020) no.2, 029901]
doi:10.1103/PhysRevLett.116.169902
[arXiv:1208.5102 [gr-qc]].

\bibitem{Yagi:2012ya}
K.~Yagi, N.~Yunes and T.~Tanaka,
Phys. Rev. D \textbf{86} (2012), 044037
[erratum: Phys. Rev. D \textbf{89} (2014), 049902]
doi:10.1103/PhysRevD.86.044037
[arXiv:1206.6130 [gr-qc]].

\bibitem{Molina:2010fb}
C.~Molina, P.~Pani, V.~Cardoso and L.~Gualtieri,
Phys. Rev. D \textbf{81} (2010), 124021
doi:10.1103/PhysRevD.81.124021
[arXiv:1004.4007 [gr-qc]].
\bibitem{Izaurieta:2009hz}
F.~Izaurieta, E.~Rodriguez, P.~Minning, P.~Salgado and A.~Perez,
Phys. Lett. B \textbf{678} (2009), 213-217
doi:10.1016/j.physletb.2009.06.017
[arXiv:0905.2187 [hep-th]].
\bibitem{Smith:2007jm}
T.~L.~Smith, A.~L.~Erickcek, R.~R.~Caldwell and M.~Kamionkowski,
Phys. Rev. D \textbf{77} (2008), 024015
doi:10.1103/PhysRevD.77.024015
[arXiv:0708.0001 [astro-ph]].

\bibitem{Konno:2009kg}
K.~Konno, T.~Matsuyama and S.~Tanda,
Prog. Theor. Phys. \textbf{122} (2009), 561-568
doi:10.1143/PTP.122.561
[arXiv:0902.4767 [gr-qc]].

\bibitem{Sopuerta:2009iy}
C.~F.~Sopuerta and N.~Yunes,
Phys. Rev. D \textbf{80} (2009), 064006
doi:10.1103/PhysRevD.80.064006
[arXiv:0904.4501 [gr-qc]].

\bibitem{Matschull:1999he}
H.~J.~Matschull,
Class. Quant. Grav. \textbf{16} (1999), 2599-2609
doi:10.1088/0264-9381/16/8/303
[arXiv:gr-qc/9903040 [gr-qc]].


\bibitem{Haghani:2017yjk}
Z.~Haghani, T.~Harko and S.~Shahidi,
Eur. Phys. J. C \textbf{77} (2017) no.8, 514
doi:10.1140/epjc/s10052-017-5078-0
[arXiv:1704.06539 [gr-qc]].






\bibitem{Kawai:2017kqt}
S.~Kawai and J.~Kim,
Phys. Lett. B \textbf{789} (2019), 145-149
doi:10.1016/j.physletb.2018.12.019
[arXiv:1702.07689 [hep-th]].




\bibitem{Satoh:2007gn}
M.~Satoh, S.~Kanno and J.~Soda,
Phys. Rev. D \textbf{77} (2008), 023526
doi:10.1103/PhysRevD.77.023526
[arXiv:0706.3585 [astro-ph]].

\bibitem{Nair:2019iur}
R.~Nair, S.~Perkins, H.~O.~Silva and N.~Yunes,
Phys. Rev. Lett. \textbf{123} (2019) no.19, 191101
doi:10.1103/PhysRevLett.123.191101
[arXiv:1905.00870 [gr-qc]].

\bibitem{Satoh:2008ck}
M.~Satoh and J.~Soda,
JCAP \textbf{09} (2008), 019
doi:10.1088/1475-7516/2008/09/019
[arXiv:0806.4594 [astro-ph]].

\bibitem{Satoh:2010ep}
M.~Satoh,
JCAP \textbf{11} (2010), 024
doi:10.1088/1475-7516/2010/11/024
[arXiv:1008.2724 [astro-ph.CO]].

\bibitem{Antoniadis:1993jc}
I.~Antoniadis, J.~Rizos and K.~Tamvakis,
Nucl. Phys. B \textbf{415} (1994), 497-514
doi:10.1016/0550-3213(94)90120-1
[arXiv:hep-th/9305025 [hep-th]].

\bibitem{GBM:2017lvd}
  B.~P.~Abbott {\it et al.}
  ``Multi-messenger Observations of a Binary Neutron Star Merger,''
  Astrophys.\ J.\  {\bf 848} (2017) no.2,  L12
  doi:10.3847/2041-8213/aa91c9
  [arXiv:1710.05833 [astro-ph.HE]].




\bibitem{Ezquiaga:2017ekz}
  J.~M.~Ezquiaga and M.~Zumalacarregui,
  Phys.\ Rev.\ Lett.\  {\bf 119} (2017) no.25,  251304
  doi:10.1103/PhysRevLett.119.251304
  [arXiv:1710.05901 [astro-ph.CO]].
  
\bibitem{Odintsov:2020sqy}
S.~D.~Odintsov, V.~K.~Oikonomou and F.~P.~Fronimos,
Nucl. Phys. B \textbf{958} (2020), 115135
doi:10.1016/j.nuclphysb.2020.115135
[arXiv:2003.13724 [gr-qc]].

\bibitem{Oikonomou:2020sij}
V.~K.~Oikonomou and F.~P.~Fronimos,
Class. Quant. Grav. \textbf{38} (2021) no.3, 035013
doi:10.1088/1361-6382/abce47
[arXiv:2006.05512 [gr-qc]].

\bibitem{Odintsov:2020xji}
S.~D.~Odintsov, V.~K.~Oikonomou and F.~P.~Fronimos,
Annals Phys. \textbf{420} (2020), 168250
doi:10.1016/j.aop.2020.168250
[arXiv:2007.02309 [gr-qc]].

\bibitem{Oikonomou:2020oil}
V.~K.~Oikonomou and F.~P.~Fronimos,
EPL \textbf{131} (2020) no.3, 30001
doi:10.1209/0295-5075/131/30001
[arXiv:2007.11915 [gr-qc]].

\bibitem{Odintsov:2020mkz}
S.~D.~Odintsov, V.~K.~Oikonomou, F.~P.~Fronimos and S.~A.~Venikoudis,
Phys. Dark Univ. \textbf{30} (2020), 100718
doi:10.1016/j.dark.2020.100718
[arXiv:2009.06113 [gr-qc]].

\bibitem{Oikonomou:2020tct}
V.~K.~Oikonomou and F.~P.~Fronimos,
Eur. Phys. J. Plus \textbf{135} (2020) no.11, 917
doi:10.1140/epjp/s13360-020-00926-3
[arXiv:2011.03828 [gr-qc]].

\bibitem{Venikoudis:2021irr}
S.~A.~Venikoudis and F.~P.~Fronimos,
Eur. Phys. J. Plus \textbf{136} (2021) no.3, 308
doi:10.1140/epjp/s13360-021-01298-y
[arXiv:2103.01875 [gr-qc]].
  
\bibitem{Hwang:2005hb}
J.~c.~Hwang and H.~Noh,
Phys. Rev. D \textbf{71} (2005), 063536
doi:10.1103/PhysRevD.71.063536
[arXiv:gr-qc/0412126 [gr-qc]].
  
\bibitem{Akrami:2018odb}
Y.~Akrami \textit{et al.} [Planck],
Astron. Astrophys. \textbf{641} (2020), A10
doi:10.1051/0004-6361/201833887
[arXiv:1807.06211 [astro-ph.CO]].
  
\bibitem{Karydas:2021wmx}
S.~Karydas, E.~Papantonopoulos and E.~N.~Saridakis,
[arXiv:2102.08450 [gr-qc]].

\bibitem{Geng:2017mic}
C.~Q.~Geng, C.~C.~Lee, M.~Sami, E.~N.~Saridakis and A.~A.~Starobinsky,
JCAP \textbf{06} (2017), 011
doi:10.1088/1475-7516/2017/06/011
[arXiv:1705.01329 [gr-qc]].

\bibitem{Pozdeeva:2021nmz}
E.~O.~Pozdeeva,
Universe \textbf{7} (2021) no.6, 181
doi:10.3390/universe7060181
[arXiv:2105.02772 [gr-qc]].

\bibitem{Pozdeeva:2021iwc}
E.~O.~Pozdeeva and Y.~Vernov,
[arXiv:2104.04995 [gr-qc]].

\bibitem{Granda:2021xyc}
L.~N.~Granda and D.~F.~Jimenez,
Eur. Phys. J. C \textbf{81} (2021) no.1, 10
doi:10.1140/epjc/s10052-020-08789-9

\bibitem{Aoki:2020ila}
K.~Aoki, M.~A.~Gorji, S.~Mizuno and S.~Mukohyama,
JCAP \textbf{01} (2021), 054
doi:10.1088/1475-7516/2021/01/054
[arXiv:2010.03973 [gr-qc]].

\bibitem{Fomin:2020hfh}
I.~Fomin,
Eur. Phys. J. C \textbf{80} (2020) no.12, 1145
doi:10.1140/epjc/s10052-020-08718-w
[arXiv:2004.08065 [gr-qc]].

\bibitem{Rashidi:2020wwg}
N.~Rashidi and K.~Nozari,
Astrophys. J. \textbf{890}, 58
doi:10.3847/1538-4357/ab6a10
[arXiv:2001.07012 [astro-ph.CO]].

\end{thebibliography}
\end{document}